\documentclass[12pt,preprint]{aastex}

\shorttitle{M Subdwarf Binaries}
\shortauthors{Riaz, Gizis & Sammaddar}

\begin{document}

\title{{\it Hubble Space Telescope} Search for M Subdwarf Binaries}

\author{Basmah Riaz\altaffilmark{1}, John E. Gizis\altaffilmark{1}, Debasmita Samaddar\altaffilmark{2}}
\affil{basmah@udel.edu, gizis@udel.edu, ftds1@uaf.edu}
\altaffiltext{1}{Department of Physics and Astronomy, University of Delaware,
    Newark, DE 19716}
\altaffiltext{2}{Physics Department, University of Alaska Fairbanks, 900 Yukon Dr., Rm 102, PO Box 755920, Fairbanks, AK 99775-5920}    
    
\begin{abstract}
We present {\it HST}/ACS observations of 19 nearby M subdwarfs in a search for binary systems. Other than the wide common proper motion pair LHS 2140/2139, none of our sdM and esdM targets are found to be binaries. Our survey is sensitive to equal-luminosity companions at close (2-8 AU) separations, while sub-stellar secondaries could have been detected at separations in the range of 6-30 AU. To check for wide binaries, we have compared the POSS I and II images in a field of view as large as 10$\arcmin$$\times$10$\arcmin$, but could not detect a single co-moving star for any of the targets. Combining our results with those from Gizis \& Reid, we have a binary fraction of 3\% (1/28). Detection of a small number of M subdwarf binaries reported in the literature suggests a higher fraction than the one obtained here, probably comparable to that found for the more massive solar-type stars in the halo (13-15\%). Comparison with the disk M dwarf fraction ($\sim$25\%) however suggests multiplicity to be rare among the lowest mass halo stars, implying the two populations formed under different initial conditions. The low binary fraction in our survey could be explained by selection biases. A decrease in multiplicity has been observed in the disk for masses below 0.1$M_{\sun}$, the peak in the disk mass function (MF). The globular cluster MF is found to peak at about 0.33$M_{\sun}$, with a decrease in the number of stars per unit mass below the peak mass. Our sample being composed of stars with masses between $\sim$0.2 and 0.085 $M_{\sun}$ suggests that a decrease in multiplicity similar to the disk may also be true for the halo stars, but perhaps below a mass of $\sim$0.3$M_{\sun}$.  A higher M subdwarf binary fraction may be obtained if the selected primaries have masses near or higher than the peak in the MF.
\end{abstract}


\section{Introduction}

A study of the old Population II binaries is important in understanding the star formation processes and dynamical evolution in the early Galaxy. Comparisons among the disk and halo binary fractions, the distributions of their period and mass ratios, along with the luminosity and mass functions (MF), help in understanding whether the two populations formed under similar environmental conditions, or if the star formation process is dependent on the metallicity of the environment. Among nearby solar-type main-sequence stars, Duquennoy \& Mayor (1991) report about 57\% of the stars to be binary or multiple systems. Fischer \& Marcy (1992) have found a lower fraction of $\sim$42\% for M dwarfs within 20 pc, with a peak in the separation distribution around 25 AU. Studies of binaries among the younger ($\la$5 Myr) pre-main sequence stars in nearby star-forming regions indicate the fraction to be twice as large as the field stars (White \& Ghez 2001). Among the very low-mass ($<$0.1 $M_{\sun}$) dwarf stars and brown dwarfs, binaries are found to be rare (10-30\%) and closely separated (a $<$ 20 AU), with a preference for equal-mass companions (Burgasser et al. 2006).

In the Galactic halo, searches such as the one by Ryan (1992) for common proper motion (CPM) companions in the NLTT catalog has resulted in finding some 25 wide binaries. A binary fraction of about 9\% at wide separations (a $>$ 25 AU) was obtained by Allen et al. (2000), after examining more than 1200 high-velocity and metal-poor stars. Other smaller surveys of metal-deficient stars indicate a slightly higher frequency of $\sim$14\% (e.g. Mart\'{\i}n \& Rebolo 1992). Earlier, studies by Carney (1983) and Stryker et al. (1985) found a binary fraction of 20-30\% for halo stars, about a factor of 2 lower than the one found for solar-metallicity stars. However, Latham et al. (2002) conducted a spectroscopic survey of high proper motion stars, and found no difference in the fraction of halo and disk binaries at close separations (a $\la$ 7 AU). A similar result was obtained at wide separations (a $\ga$ 100 AU) by Chanam$\acute{e}$ \& Gould (2004), who found no significant differences between the binary characteristics in the halo and the disk populations. These authors thus concluded that the two binary populations probably formed under similar environmental conditions. Zapatero Osorio \& Mart\'{\i}n (2004) report a binary fraction of about 9\% in wide orbits (a $\geq$ 30 AU) for their sample of metal-poor GKM stars, with no obvious discrepancy in the orbital separations of metal-depleted binaries and solar-metallicity multiple stars. Using a conservative limit of K-band flux ratios larger than 0.1, Zinnecker et al. (2004) obtained a binary frequency of 6-7\% in their visual binary survey of 164 halo stars in the solar neighborhood. Their survey is sensitive to separations in the range of 0.13$\arcsec$ and 3$\arcsec$. The fraction increases to $\sim$20\% for K-band flux ratios greater than 0.01, and is then comparable to the 15\% reported by Latham et al. (2002). The samples observed in these studies however are mainly made up of metal-poor G- and K-type stars, and lack observations of a large number of the lowest mass metal-poor stars, M subdwarfs. A few discoveries of M subdwarf binaries have been reported during the past few years (e.g. Gizis 1998, Monet et al. 2000, Gelino \& Kirkpatrick 2006). The total number of M subdwarf primaries observed however remains very small, making it impossible to conclude whether the binary fraction among these would be higher or lower than that found for the disk M dwarfs. The present halo luminosity function is found to peak at $M_{V}$$\sim$11.5, with a drop of the order of $\sim$10 observed for $M_{V}$ between $\sim$12 and 14 (Digby et al. 2003). If there exists a significant number of M subdwarf binary systems with faint very low-mass ($<$0.1 $M_{\sun}$) companions, this would result in a steeper drop at the faint end of the halo luminosity function than that observed. A high fraction of equal-mass M subdwarf binaries on the other hand would result in a broader peak, shifted towards the brighter end. The presence of unresolved M subdwarf binary systems could thus alter the observed halo luminosity function. The observed luminosity function can be converted into the corresponding MF by the application of a mass-luminosity relation. There exists no calibration of such a relation for the lowest-mass metal-poor stars, making it crucial to identify a binary star system suitable for empirical mass measurements. 

We present here results from a {\it Hubble Space Telescope} ({\it HST}) snapshot search for binary systems among nearby M subdwarfs. This supplements a previous snapshot program that only obtained nine observations (Gizis \& Reid 2000; hereafter GR). For their nine targets, GR failed to identify a single binary system and thus concluded that below 0.3 $M_{\sun}$, the halo binary fraction is less than or equal to that in the disk. We have extended their work here by obtaining {\it HST} Advanced Camera for Survey (ACS) images of 19 nearby M subdwarfs, with an aim to determine the binary fraction of such stars in the halo and to possibly identify a system which is suitable for determining accurate masses of the components. 

\section{Observations}

Our main selection criteria was to pick spectroscopically classified M subdwarfs that lie at nearby distances. We tried to select everything that was known within 50 pc and was not too bright to observe. We also added the latest spectral types. Out of the 19 targets, 14 lie within 50 pc, while the rest lie within 80 pc of the Sun. Since this was a snapshot program, only a fraction (19) of the allocated targets (35) were observed. Observations were obtained between August 2003 and August 2004, using the ACS Wide-Field Channel (WFC). This channel has a field of view of $\sim$202$\arcsec \times$ 202$\arcsec$ from 3700-11,000$\AA$. The filter F775W was used, which is similar to the SDSS {\it i} filter, with a peak at 7625$\AA$. Due to the snapshot mode, exposure times were constrained to less than 5 minutes. Table 1 lists the basic data for our targets, along with those in the GR sample. 

Photometry was performed using the various tasks under the IRAF/DAOPHOT package. A point spread function (psf) model was defined for each target using the task {\it psf}, which was then fitted to the psf stars and their neighbors using the {\it nstar} task. The fitted psf stars and their neighbors were then subtracted from the original image using the {\it substar} task. Subtracting out the psf stars made sure that no companions were hidden behind the target stars. Photometric observations were transformed into the ACS/WFC Vega-mag system defined in Bedin et al. (2005) by applying the following relation:

\begin{equation}
m_{775} = -2.5 log_{10} [DN/exptime] + Zp^{775} - Z_{Aperture},
\end{equation}

\noindent where $Zp^{775}$ is the zero-point for the F775W filter and has a value of 25.256. In the Vega-mag system, by definition, Vega has a magnitude of zero in all filters. $Z_{Aperture}$ is the aperture correction given as:

\begin{equation}
Z_{Aperture} = \Delta m_{PSF - AP(r)} + \Delta m_{AP(r) - AP(\infty)}.
\end{equation}

\noindent As discussed in Bedin et al., when performing aperture or PSF-fitting photometry, possible systematic errors due to the spatial variations of the PSF need to be addressed. Such effects are more prominent for small apertures ($<$ 5 pixels). The first term $\Delta$$m_{PSF - AP(r)}$ in Eq. (2) is to correct the PSF-fitted magnitude to the magnitude at a specified aperture, {\it r}. The PSF-fitted magnitudes for our targets were obtained for an aperture radius fixed at 2.5 pixels (0.125$\arcsec$). Using {\it r} of 0.3$\arcsec$, the $\Delta$$m_{PSF - AP(0.3\arcsec)}$ corrections were estimated in each individual image and were found to be between 0.2 and 0.25 mag. The second term $\Delta$$m_{AP(r) - AP(\infty)}$ is to link the magnitudes obtained at {\it r} = 0.3$\arcsec$ to an infinite aperture for which the zero-points have been calculated. For the F775W filter, the $\Delta$$m_{AP(0.3\arcsec) - AP(\infty)}$ correction is equal to 0.123 (Bedin et al. 2005). Furthermore, a gain correction factor of 0.008525 was added to the zero-point (Sirianni et al. 2005). The WFC can be operated at two gain settings of 1 and 2, with the default gain value being 1 $e^{-}$/ADU. Since all photometric calibration data are obtained with the default gain, a corrective coefficient is added to the zero-points to transform these calibrations to non-default gain observations. The {\it I}-band photometry including all corrections is listed in Table 1. These magnitudes correspond to those obtained in the F775W filter for our targets, while the F850LP filter for the stars in GR. The photometric errors are between 0.001 and 0.004 mag, but the calibration uncertainties are of the order of $\sim$0.02-0.03 mag (Bedin et al. 2005). 

The ACS (WFC) has a resolution of $\sim$0.05$\arcsec$/pix. The separation at which a companion can be resolved is a function of {\it $\delta$M}, the magnitude difference between the primary and the secondary. A faint companion that lies as close as 0.05$\arcsec$ would be easier to detect for a $M_{I}$=12 primary, compared to a $M_{I}$=9 one, since the faint secondary may be lost in the psf of the brighter primary. The limiting separations are thus smaller if {\it $\delta$M} is small. In order to determine our detection limits, we placed artificial secondaries at different separations (and different directions) from our targets. These were then retrieved using the IRAF {\it daofind} task. We were also able to visually identify them for {\it $\delta$M} as large as 6 using the IRAF {\it imexam} task. We have found that equal-luminosity secondaries are detectable at a separation of 2 pixels (0.1$\arcsec$), while the limiting separations for {\it $\delta$M} of 2, 4 and 6 are 0.13$\arcsec$, 0.25$\arcsec$ and 0.34$\arcsec$, respectively. Table 1 lists the observed and limiting magnitudes for our targets. If the target is a binary, then the observed $M_{I}$ is the {\it total} system luminosity, whereas $M_{I}$(lim) is the luminosity of the secondary. 

\clearpage

\begin{deluxetable}{cccccccccccccc}
\tabletypesize{\scriptsize}
\rotate
\tablecaption{Targets}
\tablewidth{0pt}
\tablehead{
\colhead{Star} & \colhead{Spectral Type} & \colhead{d (pc)} &  \colhead{Source\tablenotemark{a}} & \colhead{B\tablenotemark{b}} & \colhead{V} & \colhead{R} & \colhead{I\tablenotemark{d}} &
\colhead{$M_{I}$(obs)} & &\colhead{$M_{I}$(lim)}    \\
&&&&&&&&& \colhead{0.1$\arcsec$} & \colhead{0.13$\arcsec$} & \colhead{0.25$\arcsec$} & \colhead{0.34$\arcsec$}
}
\startdata

LHS161	&	esdM2.0	&	38.46 $\pm$ 7.1	&	$\pi$, 	RG05	&	16.30	&	14.75	&	13.74	&	12.63	&	9.7	&	10.5	&	11.9	&	13.7	&	15.7	\\
LHS1032	&	esdM4.5	&	76.8 $\pm$ 15.4	&	SpT/$M_{K}$, 	RG05	&	19.40	&	18.00	&	17.40	&	15.40	&	11.0	&	11.7	&	13.1	&	15.0	&	17.0	\\
LHS 1035	&	sdM6	&	43 $\pm$ 8.6	&	SpT/$M_{K}$, 	RG05	&	20.50	&	19.00	&		&	16.76	&	13.6	&	14.3	&	15.7	&	17.6	&	19.6	\\
LHS 1135	&	sdM6.5	&	50.6 $\pm$ 10.1	&	SpT/$M_{K}$, 	RG05	&	21.20	&	19.50	&		&	15.70	&	12.2	&	12.9	&	14.3	&	16.2	&	18.2	\\
LHS1074	&	sdM6.0	&	85.7 $\pm$ 17.1	&	SpT/$M_{K}$, 	RG05	&	20.60	&	19.00	&	18.20	&	16.07	&	11.4	&	12.2	&	13.6	&	15.4	&	17.4	\\
LHS2023	&	esdM6.0	&	73.9 $\pm$ 14.8	&	SpT/$M_{K}$, 	RG05	&	19.80	&	19.00	&	17.70	&	16.24	&	11.9	&	12.6	&	14.1	&	15.9	&	17.9	\\
LHS2140	&	sdM0.5	&	43.47 $\pm$ 7.7	&	ph $\pi$, 	DD89	&	16.55	&	15.12	&	14.11	&	12.90	&	9.7	&	10.5	&	11.9	&	13.7	&	15.7	\\
LHS375	&	esdM4.0	&	25.64 $\pm$ 0.65	&	$\pi$, 	RG05	&	17.55	&	15.68	&	14.60	&	13.24	&	11.2	&	11.9	&	13.4	&	15.2	&	17.2	\\
LHS3181	&	sdM2.0	&	38.46 $\pm$ 4.4	&	$\pi$, 	RA93	&	18.87	&	17.18	&	16.16	&	14.94	&	12.0	&	12.8	&	14.2	&	16.0	&	18.0	\\
LHS491	&	sdM1.5	&	47.39 $\pm$ 8.1	&	$\pi$, 	RG05	&	16.38	&	14.70	&	13.72	&	12.54	&	9.2	&	9.9	&	11.3	&	13.2	&	15.2	\\
APMPMJ2126-4454	&	sdM0	&	44 $\pm$ 8.8	&	Spec, 	S02	&	16.30	&	15.50	&	14.28	&	12.87	&	9.7	&	10.4	&	11.8	&	13.7	&	15.7	\\
LSR0014+6546	&	sdM4.5	&	40 $\pm$ 20	&	Spec, 	L03	&	18.00	&	16.50	&	15.70	&	14.16	&	11.2	&	11.9	&	13.3	&	15.2	&	17.2	\\
LSR0157+5308	&	sdM3.5	&	45 $\pm$ 22	&	Spec, 	L03	&	17.10	&	16.00	&	14.90	&	13.29	&	10.0	&	10.8	&	12.2	&	14.1	&	16.0	\\
LSR0519+4213	&	esdM3.5	&	55 $\pm$ 27	&	Spec, 	L03	&	18.80	&	17.00	&	16.10	&	14.51	&	10.8	&	11.6	&	13.0	&	14.8	&	16.8	\\
LSR0522+3814	&	esdM3.0	&	45 $\pm$ 22	&	Spec, 	L03	&	16.60	&	15.50	&	14.50	&	14.11	&	10.8	&	11.6	&	13.0	&	14.9	&	16.8	\\
LSR0609+2319	&	sdM5.0	&	45 $\pm$ 22	&	Spec, 	L03	&	18.70	&	17.50	&	16.50	&	14.49	&	11.2	&	12.0	&	13.4	&	15.2	&	17.2	\\
LSR1425+7102	&	sdM8.0	&	65 $\pm$ 15 	&	Spec, 	L03	&	20.80	&	19.50	&	18.60	&	16.31	&	12.2	&	13.0	&	14.4	&	16.3	&	18.3	\\
LSR2036+5059	&	sdM7.5	&	18 $\pm$ 9	&	Spec, 	L03	&		&	18.00	&	16.80	&	15.86	&	14.6	&	15.3	&	16.7	&	18.6	&	20.6	\\
LSR2122+3656	&	esdM5.0	&	45 $\pm$ 22	&	Spec, 	L03	&	18.70	&	17.50	&	16.20	&	14.88	&	11.6	&	12.4	&	13.8	&	15.6	&	17.6	\\
LHS169\tablenotemark{c} 	&	esdK7	&	32.4 $\pm$ 2.4	&	$\pi$, 	GR	&	15.58	&	14.13	&	13.22	&	12.15	&	9.6	&	10.3	&	11.8	&	13.6	&	15.6	\\
LHS174\tablenotemark{c} 	&	sdM0.5	&	49.01 $\pm$ 8.8	&	$\pi$, 	GR	&	14.27	&	12.75	&	11.81	&	10.57	&	7.1	&	7.9	&	9.3	&	11.1	&	13.1	\\
LHS216\tablenotemark{c} 	&	sdM2.0	&	32.7 $\pm$ 3.2	&	$\pi$, 	GR	&	16.28	&	14.66	&	13.67	&	12.13	&	9.6	&	10.3	&	11.7	&	13.6	&	15.6	\\
LHS320\tablenotemark{c} 	&	sdM2.0	&	38.5 $\pm$ 5	&	$\pi$, 	GR	&	15.54	&	14.00	&	13.12	&	11.46	&	8.5	&	9.3	&	10.7	&	12.6	&	14.5	\\
LHS3409\tablenotemark{c} 	&	sdM4.5	&	20 $\pm$ 0.52	&	$\pi$, 	GR	&	16.96	&	15.16	&	15.40	&	11.89	&	10.4	&	11.1	&	12.5	&	14.4	&	16.4	\\
LHS364\tablenotemark{c} 	&	esdM1.5	&	26.7 $\pm$ 2.64	&	$\pi$, 	GR	&	16.32	&	14.61	&	13.58	&	12.21	&	10.1	&	10.8	&	12.2	&	14.1	&	16.1	\\
LHS377\tablenotemark{c} 	&	sdM7.0	&	35.2 $\pm$ 1	&	$\pi$, 	GR	&	20.20	&	18.39	&	17.10	&	14.16	&	11.4	&	12.2	&	13.6	&	15.5	&	17.4	\\
LHS407\tablenotemark{c} 	&	esdM5.0	&	31.7 $\pm$ 2	&	$\pi$, 	GR	&	18.50	&	16.57	&	15.51	&	13.65	&	11.1	&	11.9	&	13.3	&	15.2	&	17.1	\\
LHS522\tablenotemark{c} 	&	esdK7	&	37.3 $\pm$ 3	&	$\pi$, 	GR	&	15.56	&	14.15	&	13.31	&	12.26	&	9.4	&	10.2	&	11.6	&	13.4	&	15.4	\\
LHS536\tablenotemark{c} 	&	sdM0.5	&	44.1 $\pm$ 4.85	&	$\pi$, 	GR	&	16.20	&	14.65	&	14.70	&	12.45	&	9.2	&	10.0	&	11.4	&	13.3	&	15.2	\\
LHS541\tablenotemark{c} 	&	sdM3.0	&	80.6 $\pm$ 8	&	$\pi$, 	GR	&	19.29	&	16.46	&	14.80	&	13.88	&	9.4	&	10.1	&	11.5	&	13.4	&	15.4	\\

\enddata

\tablenotetext{a}{Distance estimated from: $\pi$--trigonometric parallax; Spt/$M_{K}$-- spectral type- $M_{K}$ relation; ph $\pi$--photometric parallax; Spec--spectroscopic. References for these estimates are: RG05--Reid \& Gizis (2005);  DD89--Dawson P.C. \& De Robertis M.M. (1989); RA93--Ruiz M.T. \& Anguita C. (1993); S02--Scholz et al. (2002); L03--L\'epine et al. (2003b).}
\tablenotetext{b}{BVR magnitudes from Gizis (1997) and GR.}
\tablenotetext{c}{Targets from GR.}
\tablenotetext{d}{The {\it I}-band photometry corresponds to that obtained in the F775W filter for our targets, while the F850LP filter for the stars in GR. }

\end{deluxetable}
\clearpage

\section{Discussion}

Apart from the known white dwarf-red subdwarf system LHS 2140/2139 (Fig. 1a), we have found one other probable wide binary system in our sample. Fig. 1b shows the {\it HST}/ACS image for LHS 1074. The $m_{775}$ magnitude for this star is 16.4. The faint object with $m_{775}$ of 22.6 mag lies at a separation of $\sim$2.1$\arcsec$ ($\sim$180 AU at 85.7 pc) from LHS 1074, and is not detectable in the first and second Palomar Sky Survey (POSS I and II) images. Therefore, at present, we are unable to confirm this object to be a CPM companion. As discussed later in this section, there is a 25\% chance for this faint object to be a background star. We therefore treat this system with caution and do not include it in the discussion below. Further observations have been requested that should allow us to confirm the binarity of LHS 1074. For the case of LHS 2140/2139, Gizis \& Reid (1997) have classified LHS 2140 to be an sdM0.5 and the fainter LHS 2139 to be an old halo white dwarf. A comparison of the POSS I and II images confirms LHS 2140/2139 to be a wide CPM binary system, with a separation of about 6.7$\arcsec$ ($\sim$290 AU at 43.5 pc). A search in the literature indicates the presence of a few other such wide binaries among metal-poor M-type stars.  In their CCD-based imaging search for wide low-mass companions to 473 GKM stars with known low metallicities, Zapatero Osorio \& Mart\'{\i}n (2004) have found three M-type primaries, G 063-036, G 197-049 and G 252-049 to have visual, CPM companions at separations of 37, 225 and 32 AU, respectively. Chanam$\acute{e}$ \& Gould (2004) conducted a survey of wide (a $\ga$ 100 AU) CPM binaries, selected from the revised NLTT catalog. They have found 116 binaries in the halo, with most having spectral types between M0 and M5. The spectral types have been determined  based on their location in the V-J vs. J-K color-color diagram. Our observations are sensitive to companions at separations as close as $\sim$0.1$\arcsec$. At wide separations, we would have considered a star lying within 3$\arcsec$ of our target as a possible companion, as in the case of LHS 1074. We have compared the POSS I and II images in a field of view as large as 10$\arcmin$$\times$10$\arcmin$ to search for any co-moving stars, but none could be found for any of the other targets. 

The masses for our sample can be estimated from their absolute {\it I} magnitudes, using the evolutionary models by Barraffe et al. (1997) for metal-poor low-mass stars. For $M_{I}$ between $\sim$9 and 14, the corresponding masses are between $\sim$0.2 and 0.083 $M_{\sun}$, close to the sub-stellar borderline. In their calculation of the MF for a dozen globular clusters (GC), Paresce \& De Marchi (2000) have found the peak mass to be 0.33$\pm$0.03 $M_{\sun}$, and a decrease in the number of stars per unit mass with decreasing masses below the peak. In their multiplicity analysis of stars within 8 pc of the Sun, 80\% of which are M dwarfs, Reid \& Gizis (1997) have derived a MF $\Psi$(m) $\propto$ $M^{-1.05}$ for the mass range 1 to 0.1 $M_{\sun}$, and have found a sharp decline in number densities at masses below 0.1 $M_{\sun}$. For binaries among very-low mass ($<$0.1 $M_{\sun}$) dwarfs and brown dwarfs in the field, a decline has been seen for separations greater than 20 AU (Burgasser et al. 2006). One reason for the lack of wide binaries in our survey could be that just as wide binaries disappear for disk primaries with M $<$ 0.1 $M_{\sun}$, this could be happening as well for the halo binaries, but perhaps at a higher mass of $\sim$0.3 $M_{\sun}$. The three M dwarf primaries with CPM companions in Zapatero Osorio \& Mart\'{\i}n (2004) survey have $M_{V}$ between 7.8 and 10.7, corresponding to masses $\ga$0.3$M_{\sun}$. For the M-type halo binaries in Chanam$\acute{e}$ \& Gould (2004), the V-J colors are between $\sim$1 and 3, corresponding to M $\ga$ 0.2$M_{\sun}$.  A higher M subdwarf binary fraction may be obtained if the selected primaries have masses near or higher than the peak in the GC MF. We note that the halo MF may be different than that derived for the GCs. For the Galactic halo, Paresce \& De Marchi (2000) have found that the MF obtained from the luminosity function of Dahn et al. (1995) and corrected for binaries by Graff \& Freese (1996) peaks at about 0.16 $M_{\sun}$, and then drops for lower masses, down to the sub-stellar limit. There is also a steeper rise in number density between $\sim$0.8 and 0.25 $M_{\sun}$ than the disk MF. The halo MF, however, is known with low accuracy since the total sample of field subdwarfs observed is too small, and the available data has poorly-defined abundances. Field subdwarfs span a large range in abundances. With decreasing metallicity, the (mass, $M_{V}$) relationship predicted by theoretical models indicates brighter absolute V magnitude for a given mass. This effect is less pronounced in the predicted (mass, $M_{K}$) relationship, as the atomic and molecular features are more prominent at optical wavelengths. But since few field subdwarfs have infrared photometry, the (mass, $M_{V}$) relationship is used to derive the halo MF, which results in uncertainties of 20-30\% in the inferred mass. The halo MF can also be estimated from GC, since the abundances for the latter are known with relatively high precision, and there is no uncertainty involved in the appropriate mass-luminosity relationship (Reid \& Hawley 2000). For our case, even if we consider the halo MF, our targets lie in the declining region of the MF below the peak mass, suggesting a lower probability of detecting binaries among these.

Gelino \& Kirkpatrick (2006) have recently detected an extreme M subdwarf (esdM) wide binary system at a separation of 36 AU. These authors also report the detection of a close M subdwarf binary at a separation of 0.15$\arcsec$ (2.7 AU at 18 pc). A few other M subdwarf binaries have been reported in the literature: LSR 1530+5608 is a common proper motion near-equal mass esdM system (Monet et al. 2000, L\'epine et al. 2003a), Gl 455 is a near-equal mass sdM spectroscopic binary (Gizis 1998), while Gl 781 (Gizis 1998), LP 164-51/52 (Silvestri et al. 2002), LHS 193 and LHS 300 (Jao et al. 2005) are white dwarf-red subdwarf systems. Table 1 indicates that we could have detected equal-luminosity companions at separations in the range of 2-8 AU. Zinnecker et al. (2004) have found the projected separation distribution for their G- and K-type halo binaries to exhibit a bimodality, with peaks at $\sim$10 and $\sim$500 AU. If this is true for the lowest-mass halo members too, then it may be that we are looking at the dip in this distribution at about 60 AU, where the number of companions per system after corrections for chance alignment is found to be only 0.04, compared to 0.15 at the peaks (Zinnecker et al. 2004). We note that the reality of this bimodality is not firmly established, and their distribution is based on small numbers. Earlier work by Abt \& Levy (1969) had found the fraction of short-period binaries to be lower among the high-velocity stars, compared to the low-velocity ones. However, Stryker et al. (1985) calculations suggest that the initial binary-system separation distribution has not altered significantly with time. Indeed, Latham et al. (2002) have found similar period distributions for the disk and halo binaries at separations less than 7 AU, suggesting that metallicity has little effect on the fragmentation process that leads to short-period binaries.  The observations presented here are not sensitive enough to detect very short-period binaries (separations less than $\sim$0.1$\arcsec$), and these could have been missed out in our survey. K$\ddot{o}$hler et al. (2001) have found the distribution of mass ratios for Population II binaries to rise towards smaller values, in comparison with that for the Population I stars that peaks at about 0.3 (Duquennoy \& Mayor 1991). Our survey is sensitive to sub-stellar companions at separations in the range of 6-30 AU. However, none could be detected for any of the targets. 

Combining our results with those from GR, we have a binary fraction of about 3\% (1/28). Including the probable wide binary system LHS 1074 increases the fraction to 7\%. Reid \& Gizis (1997) report a fraction of 35\% for disk M dwarfs within 8 pc of the Sun. The multiplicity fraction (1/28) measured here with the ACS data is the fraction of stars which have companions at separations larger than $\sim$6 AU. The measured binary fraction for systems with separations $>$6 AU in Reid \& Gizis (1997) is about 25\%. For low-mass binaries in the Hyades, Gizis \& Reid (1995) report a fraction of 20\%$\pm$10\% (projected separations $>$6 AU). Using the method described in Burgasser et al. (2003), we have constructed a probability distribution for the binary fraction, $\epsilon_{b}$, for a sample size of {\it N} = 28 and number of binaries {\it n} = 1 (Fig. 2). The shaded region in Fig. 2 is one standard deviation from the mean, and has an integrated probability of 68.27\%. In comparison with the disk M dwarf fraction, the probability that the M subdwarf binary fraction is $\ga$20\% is found to be only $\sim$3\%, at a confidence level of 99.73\%. This indicates a weak probability (at a high confidence level) for the M subdwarf fraction to be as large as what is found in the disk. On the other hand, the presence of a few M subdwarf binary systems mentioned above suggests that the fraction must be higher than the 3\% obtained here. Gelino \& Kirkpatrick (2006) have reported a 13\% fraction for metal-poor GKM stars, similar to the 15\% obtained by Latham et al. (2002) and Zinnecker et al. (2004) for GK subdwarfs. Fig. 2 indicates that the probability for the M subdwarf binary fraction to be between 10 and 15\% is about 17\%, at a confidence level of 68.3\%. Thus while the binary frequency for the lowest-mass stars in the halo may be less than that in the disk, there is a fair probability that it may be as high as what is found for the solar-type halo stars. Among these higher mass stars, while no significant differences have been observed by Latham et al. (2002), Zapatero Osorio \& Mart\'{\i}n (2004) and other surveys in the characteristics of the disk and halo binaries, the fractions reported by these authors (13-15\%) are much smaller than that found by Duquennoy \& Mayor (1991; 57\%). If we consider only the wide-binary fraction with separations $\ga$30 pc, then the fraction reported by Duquennoy \& Mayor would be half of 57\%, or about 30\%. Nevertheless, this suggests that overall, multiplicity may be rarer among GKM halo stars than those in the disk population. As suggested by Stryker et al. (1985), since the initial distribution of binary-system separation has not been greatly altered with time, any differences in the binary frequency of the disk and halo populations would reflect differing conditions in the collapsing Galaxy and its effects on stellar condensation. 

Considering the case of LHS 1074, we can estimate the expected number of stars within 3$\arcsec$ of the target. The ACS (WFC) channel has a field of view of $\sim$202$\arcsec \times$ 202$\arcsec$. Within this field of view, the estimated number of background stars is found to be 3.5E-04, which results in the expected number of stars within 3$\arcsec$ to be 0.0099. In other words, there is a $\sim$1\% probability of detecting a background star within a 3$\arcsec$ radius of the target, or a 99\% probability of not detecting any background star within this radius. In a sample size of 28 then, there is a 3 in 4 chance of not finding a background star within 3$\arcsec$. The faint object at 2.1$\arcsec$ separation from LHS 1074 thus has a weak 25\% probability of being a background star. If indeed LHS 1074 is a binary and this is not a chance alignment, then an absolute {\it I} magnitude of 17.93 would make the faint star most likely sub-stellar. 

Dynamical evolution can lead to significant differences between the present day MF and the initial MF (IMF). The presence of a substantial number of unresolved binaries may drastically alter the present day MF. The luminosity of a binary is caused by light from both stars. If the light is interpreted to be only from a single star, then the mass of the star would be overestimated. Graff \& Freese (1996) have studied the effects of unresolved binaries on the derived MF in the halo, and have shown that the MFs generated by their binary models are steeper towards lower masses compared to the no-binary model. Decreasing the metallicity also causes the MF to steepen towards lower masses. De Marchi et al. (2004) have discussed whether the shift seen in the peak of the IMF for GCs, compared to that for young clusters (YC), indicates dynamical evolution in the GCs. For YCs of ages between $\sim$0.5 and $\sim$800 Myr, these authors estimate the IMF to peak at $\sim$0.15 $M_{\sun}$. The present day GC MF is found to peak at about 0.33 $M_{\sun}$. Since the MF of a dozen GCs are found to be very similar (Paresce \& De Marchi 2000), this suggests they could all share the same IMF with a peak at $\sim$0.33 $M_{\sun}$. As discussed in De Marchi et al. (2004), since any low-mass stars lost by GCs should populate the halo, if the IMF of GCs was originally similar to that of the YCs, then the halo MF should also peak at $\sim$0.15 $M_{\sun}$. This is found to be true. However, the large uncertainties due to various factors discussed earlier make the current derivation of the halo MF inaccurate. On the other hand, if the presence of a substantial number of unresolved binaries results in a halo MF similar to that currently observed in the GCs, this would indicate that their present day MF does not differ significantly from the IMF (De Marchi et al. 2004). Observations of large samples of metal-poor low-mass stars are required to accurately determine the halo binary fraction and test these hypotheses. Our survey is not sensitive enough to detect very short-period binaries (separations less than $\sim$0.1$\arcsec$), or sub-stellar companions at separations less than 6 AU. Radial velocity surveys can play an important role in detecting sub-stellar companions or objects at very close separations, that may have been missed out in imaging searches.

\section{Summary}
The absence of any binary systems in our sample other than LHS 2140/2139 could be explained by selection biases. Among disk stars, a drop in the wide ($>$ 20 AU) binary frequency is observed for primaries with masses less than 0.1$M_{\sun}$. This is consistent with the sharp decline observed in the number of stars with masses below the peak (0.1 $M_{\sun}$) in the disk MF. The GC MF is found to peak at about 0.33$M_{\sun}$, with a drop in the number of stars per unit mass below the peak mass. All of our targets have masses below $\sim$0.2 $M_{\sun}$, suggesting that the decrease in multiplicity with decreasing mass seen in the disk for masses below 0.1$M_{\sun}$ might also be happening in the halo, but at a higher mass of about 0.3$M_{\sun}$. A higher fraction may be obtained if the M subdwarf primaries selected have masses higher than the peak in the MF. Combining ours and GR survey, a sample of 28 is still too small in size. A larger sample of metal-poor M-type stars needs to be examined to determine the differences in the binary frequencies of the disk and the halo populations, that may reflect upon the differing conditions in the early Galaxy and its effects on binary formation.

\acknowledgments
This work is based on observations made with the NASA/ESA Hubble Space Telescope, obtained at the Space Telescope Science Institute, which is operated by the Association of Universities for Research in Astronomy, Inc., under NASA contract NAS 5-26555. These observations are associated with program $\#$ 9842. Support for program $\#$ 9842 was provided by NASA through a grant from the Space Telescope Science Institute, which is operated by the Association of Universities for Research in Astronomy, Inc., under NASA contract NAS 5-26555. This work has made use of the SIMBAD database.

{\it Facilities:} \facility{Hubble Space Telescope}

\clearpage
\begin{figure}
 \begin{center}
    \begin{tabular}{cc}
      \resizebox{71mm}{!}{\includegraphics[angle=0]{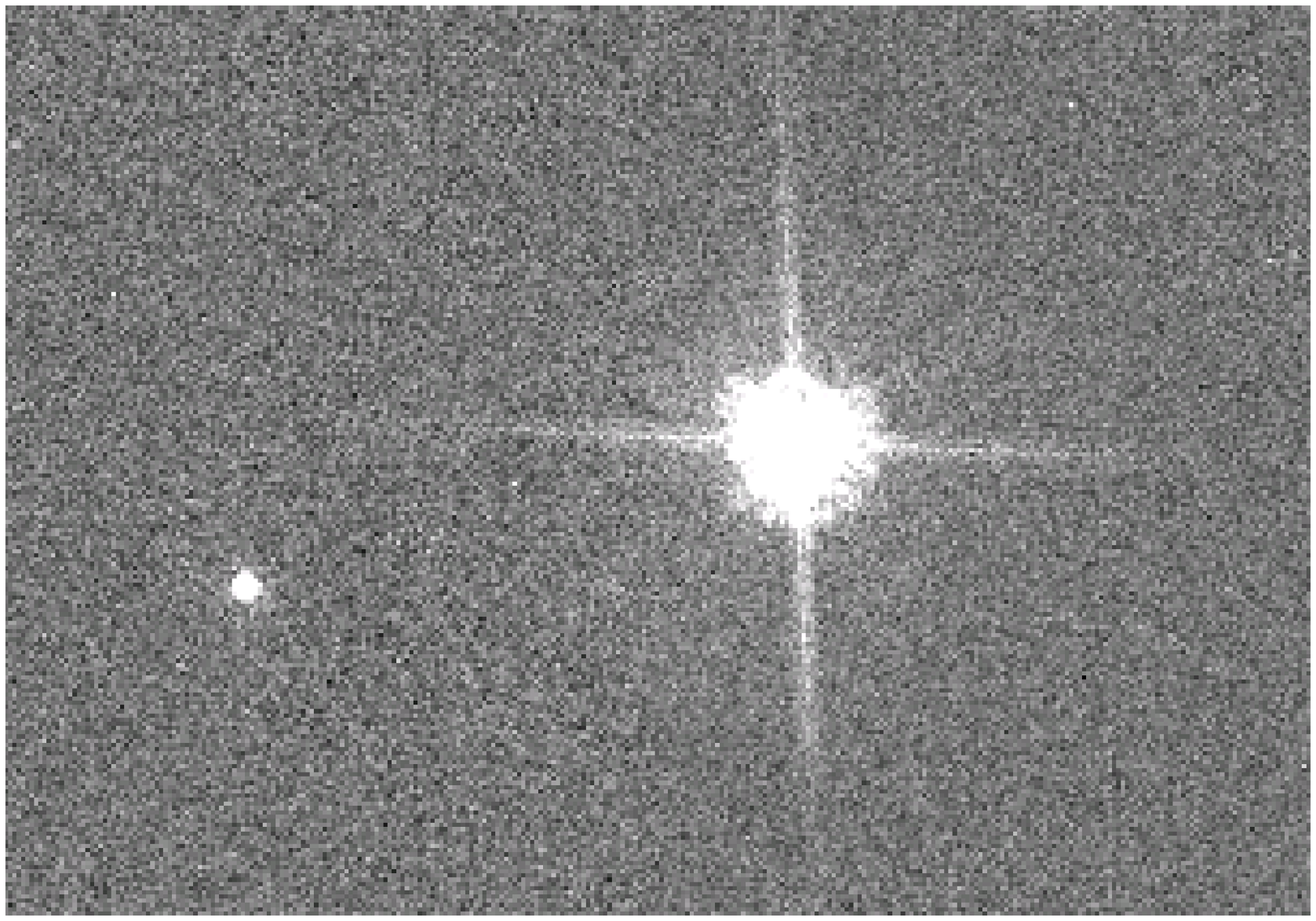}} &
      \resizebox{70mm}{!}{\includegraphics[angle=0]{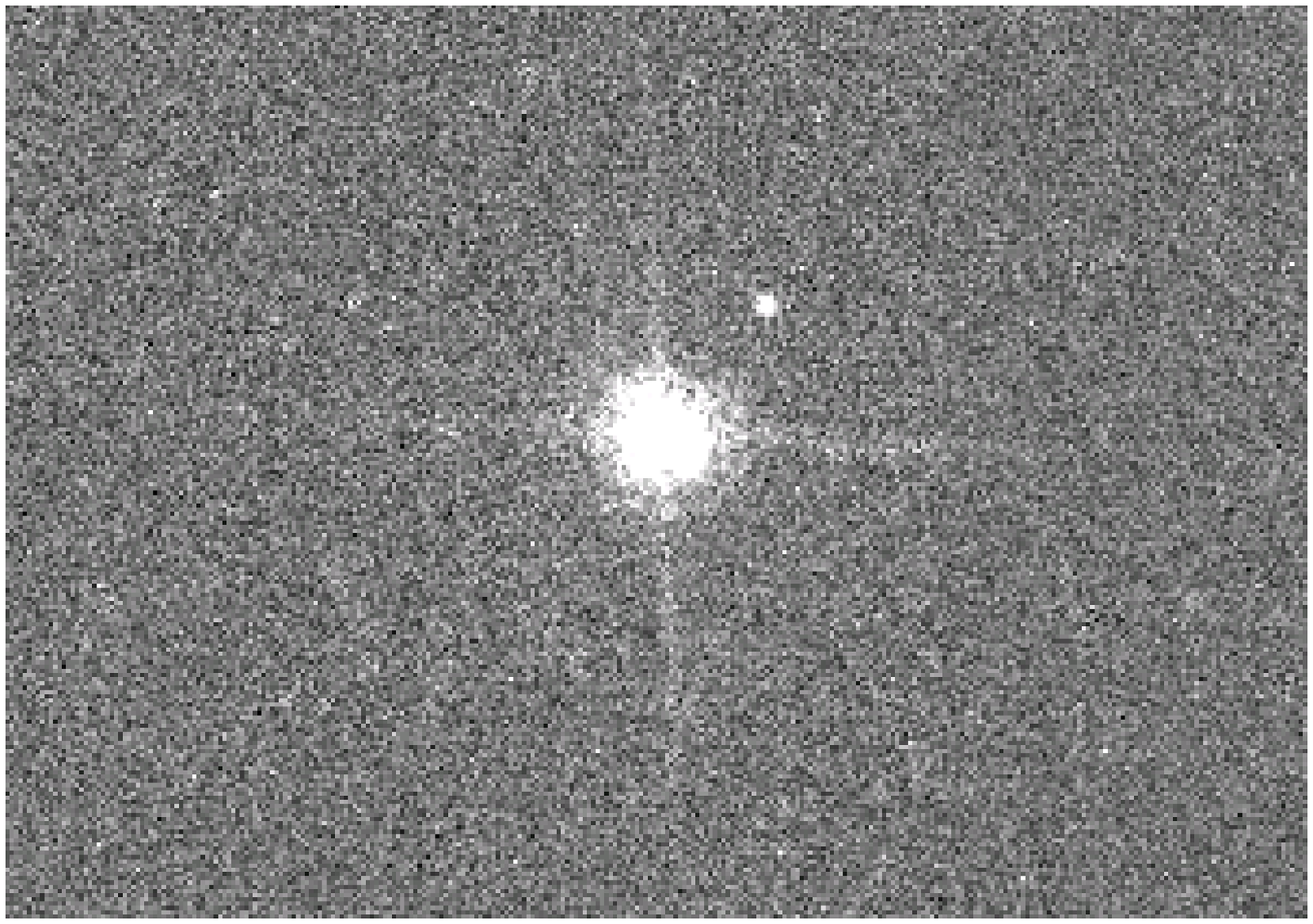}} \\
    \end{tabular}
   \caption{{\it HST}/ACS images for {\it left}- (a) LHS 2140/2139 and {\it right}- (b) LHS 1074. The faint object with $m_{775}$ = 22.6 lies at a separation of $\sim$2.1$\arcsec$ from LHS 1074.}
  \end{center}
\end{figure}

\begin{figure}
\resizebox{100mm}{!}{\includegraphics[angle=0]{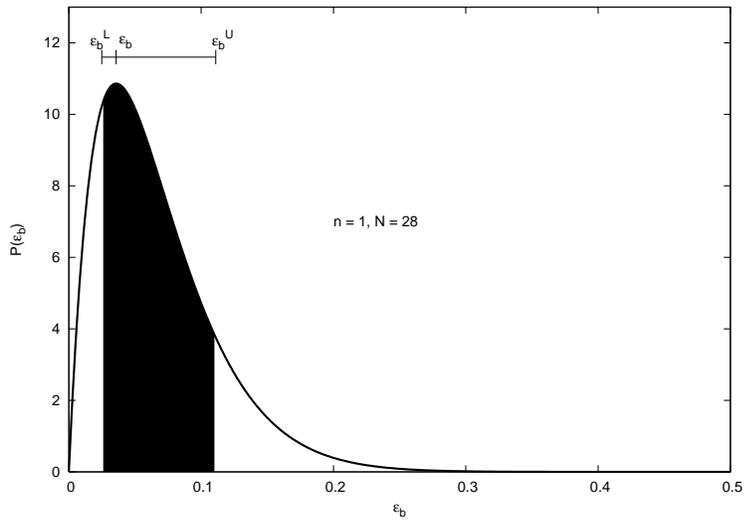}}
\caption{Probability distribution for $\epsilon_{b}$, the binary fraction, constructed for a sample size {\it N} = 28 and number of binaries {\it n} = 1. The shaded region has an integrated probability of 68.3\%, and is equivalent to 1-$\sigma$ Gaussian limits. }
\end{figure}


\begin{thebibliography}{0}
\bibitem[Abt \& Levy (1969)]{al69} Abt, H. A. \& Levy, S. G., 1969, 
\bibitem[Allen et al. (2000)]{a00} Allen, C., Poveda, A., \& Herrera, M. A. 2000, \aap, 356, 529
\bibitem[Baraffe et al. (1997)]{b97} Baraffe, I. et al. 1997, \aap, 327, 1054
\bibitem[Bedin et al. (2005)]{bed05} Bedine, L. R., et al. 2005, MNRAS, 357, 1038
\bibitem[Burgasser et al. (2003)]{b03} Burgasser, A. J., et al. 2003, \apj, 586, 512
\bibitem[Burgasser et al. (2006)]{b06} Burgasser, A. J., et al. 2006, astro-ph/0602122
\bibitem[Carney (1983)]{c83} Carney, B. 1983, \aj, 101, 623
\bibitem[Chanam$\acute{e}$ \& Gould (2004)]{cg04} Chanam$\acute{e}$, J. \& Gould, A. 2004, \apj, 601, 289
\bibitem[Dahn et al. (1995)]{dahn95} Dahn, C., et al.  1995, in ESO Conf. Proc., The Bottom of the Main-Sequence and Beyond, ed. C. G. Tinney (Berlin: Springer), 239
\bibitem[Dawson \& De Robertis (1989)]{dd89} Dawson, P. C. \& De Robertis M. M. 1989, \aj, 98, 1472
\bibitem[De Marchi et al (2004)]{de04} De Marchi, G. et al. 2004, astro-ph/0409601
\bibitem[Delfosse et al. (2004)]{del04} Delfosse, X., et al. 2004, ASP Conf. Ser., 318, 166
\bibitem[Digby et al. (2003)]{dig03} Digby, A.P. et al. 2003, MNRAS, 344, 583
\bibitem[Duquennoy \& Mayor (1991)]{dm91} Duquennoy, G. \& Mayor, M. 1991, \aap, 248, 485
\bibitem[Fischer \& Marcy (1992)]{fm92} Fischer, D. A. \& Marcy, G. W. 1992, \apj, 396, 178
\bibitem[Gelino \& Kirkpatrick (2006)]{gk06} Gelino, C. R. \& Kirkpatrick, J. D. 2006, 14th Cambridge Workshop on Cool Stars, Stellar Systems, and the Sun; Abstract$\#$148
\bibitem[Graff \& Freese (1996)]{gf96} Graff, D. S. \& Freese, K., 1996, \apj, 467, L65
\bibitem[Gizis \& Reid (1995)]{gr95} Gizis, J. E., \& Reid, I. N. 1995, \aj, 110, 3
\bibitem[Gizis (1997)]{g97} Gizis, J. E. 1997, \aj, 113, 806
\bibitem[Gizis \& Reid (1997)]{GR97} Gizis, J. E., \& Reid, I. N. 1997, PASP, 109, 849
\bibitem[Gizis (1998)]{g98} Gizis, J. E. 1998, \aj, 115, 2053
\bibitem[Gizis \& Reid (2000)]{GR00} Gizis, J. E., \& Reid, I. N. 2000, PASP, 112, 610 (GR)
\bibitem[Jao et al. (2005)]{j05} Jao, W. -C., et al., 2005, \aj, 129, 1954
\bibitem[K$\ddot{o}$hler et al. (2001)]{k01} K$\ddot{o}$hler, R., Zinnecker, H. \& Jahrei$\beta$, H. 2001, ASP Conference Series, Vol. 228, 491
\bibitem[Latham et al. (2002)]{l02} Latham, D. W., et al., 2002, \aj, 124, 1144
\bibitem[L\'epine et al. (2003a)]{l03a} L\'epine, S., Shara, M.  M., Rich, R. M. 2003, \aj, 126, 921
\bibitem[L\'epine et al. (2003b)]{l03b} L\'epine, S., Rich, R. M., Shara, M.  M. 2003, \aj, 125, 1598
\bibitem[Mart\'{\i}n \& Rebolo (1992)]{mar92} Mart\'{\i}n, E. L., \& Rebolo, R. 1992, ASP Conf. Ser., 32, 336
\bibitem[Monet et al. (2000)]{m00} Monet, D. G., et al., 2000, \aj, 120, 1541
\bibitem[Paresce \& De Marchi (2000)]{pde00} Paresce, F. \& De Marchi, G. 2000, \apj, 534, 870
\bibitem[Reid \& Gizis (1997)] {rg97} Reid, I. N. \& Gizis, J. E., 1997, \aj, 113, 2246
\bibitem[Reid \& Gizis (2005)] {rg05} Reid, I. N. \& Gizis, J. E., 2005, PASP, 117, 676
\bibitem[Reid \& Hawley]{rh00} Reid, I. N. \& Hawley, S. L., 2000, New Light on Dark Stars: Red Dwarfs, Low-Mass Stars, Brown Dwarfs (New York, Springer-Praxis)
\bibitem[Reid et al. (2003)]{r03} Reid, I. N., et al., 2003, \aj, 125, 354
\bibitem[Ruiz \& Anguita (1993)]{rz93} Ruiz M.T. \& Anguita C. 1993, \aj, 105, 614
\bibitem[Ryan (1992)]{r92} Ryan, S. G., 1992, \aj, 104, 1144
\bibitem[Scholz et al. (2002)]{sc02} Scholz R.-D. et al., 2002, MNRAS, 329, 109
\bibitem[Silvestri et al. (2002)]{s02} Silvestri, N. M., Oswalt, T. D., \& Hawley, S. L. 2002, \aj, 124, 1118
\bibitem[Sirianni et al. (2005)]{sir05} Sirianni, M., et al., 2005, PASP, 117, 1049
\bibitem[Stryker et al. (1985)]{s85} Stryker, L. L., et al. 1985, PASP, 97, 247
\bibitem[White \& Ghez (2001)]{wg01} White, R. J. \& Ghez, A. M. 2001, \apj, 556, 265
\bibitem[Zapatero Osorio \& Mart\'{\i}n (2004)]{zom04} Zapatero Osorio, M. R. \& Mart\'{\i}n, E. L., 2004, \aap, 419, 167
\bibitem[Zinnecker et al. (2004)]{zinn04} Zinnecker, H., K$\ddot{o}$hler, R. \& Jahrei$\beta$, H. 2004, RevMexAA, 21, 33

\end{thebibliography}
\end{document}